\documentclass[twocolumn,prl, aps,superscriptaddress,showpacs]{revtex4}

\usepackage{mathptmx}
\usepackage{subfigure}
\usepackage{dcolumn}
\usepackage{amsmath,amssymb}
\usepackage{bm}
\usepackage{color}
\usepackage{latexsym}
\usepackage{epstopdf}
\usepackage{color}
\usepackage[english]{babel}
\usepackage{latexsym}

\usepackage{psfrag,graphicx} 
\usepackage{epsf} 
\usepackage{subfigure} 
\usepackage{amsmath} 
\usepackage{amssymb} 
\usepackage{amsfonts}
\usepackage{bm}
\usepackage{natbib}
\usepackage{epstopdf}
\usepackage{appendix}

\definecolor{mygrey}{gray}{0.35}
\definecolor{myblue}{rgb}{0.,0.,1}
\definecolor{myzard}{cmyk}{0,0,0.05,0}
\definecolor{mywhite}{rgb}{1,1,1}
\definecolor{myred}{rgb}{1,0.,0.3}

\usepackage[colorlinks=true,citecolor=blue,linkcolor=blue]{hyperref}

\def\be{\begin{equation}}
\def\ee{\end{equation}}
\def\ba{\begin{align}}
\def\enda{\end{align}}
\def\bi{\begin{itemize}}
\def\ei{\end{itemize}}

 \def\ee{\mathord{\rm e}}
 
 \def\ii{\mathord{\rm i}}

 \def\ee{\mathord{\rm e}}
 
 \def\ii{\mathord{\rm i}}

\renewcommand{\ii}{{\rm i}}
\renewcommand{\ee}{{\rm e}}

\def\beq{\begin{equation}}
\def\beq{\begin{equation}}
\def\eeq{\end{equation}}

 \newcommand{\ket}[1]{|#1\rangle}
 \newcommand{\bra}[1]{\langle #1|}

\begin{document}

\pacs{37.10.Jk, 67.85.-d,11.15.Ha,71.10.Fd}

\title{Wilson Fermions and Axion Electrodynamics in Optical Lattices}
\author{A. Bermudez 
}
\affiliation{
Departamento de F\'isica Te\'orica I,
Universidad Complutense, 
28040 Madrid, 
Spain
}

\author{L. Mazza}
\affiliation{
Max-Planck-Institut f\"{u}r Quantenoptik, Hans-Kopfermann-Strasse 1, 
85748 
Garching, Germany
}

\author{M. Rizzi}
\affiliation{
Max-Planck-Institut f\"{u}r Quantenoptik, Hans-Kopfermann-Strasse 1, 
85748 
Garching, Germany
}

\author{
N. Goldman
}
\affiliation{Center for Nonlinear Phenomena and Complex Systems - Universit$\acute{e}$ Libre de Bruxelles , 231, Campus Plaine, B-1050 Brussels, Belgium}

\author{ 
M. Lewenstein}

\affiliation{
ICFO-Institut de Ci\`encies Fot\`oniques,
Parc Mediterrani de la Tecnologia,
E-08860 Castelldefels (Barcelona), Spain}
\affiliation{
ICREA - Instituci\'o Catalana de Recerca i Estudis Avan{\c c}ats, 08010 
Barcelona, Spain}

\author{M.A. Martin-Delgado}

\affiliation{
Departamento de F\'isica Te\'orica I,
Universidad Complutense, 
28040 Madrid, 
Spain
}

\begin{abstract}
We show that ultracold Fermi gases in optical superlattices can be used as quantum simulators of different types of relativistic lattice fermions in 3+1 dimensions. By exploiting laser-assisted tunneling methods, we find the atomic analogue of the so-called {\it naive Dirac fermions},  and thus provide a physical realization of the fermion doubling problem. Moreover, we show how to implement {\it Wilson fermions}, and  discuss how their mass can be inverted by tuning the laser intensities. In this regime, our atomic gas corresponds to a remarkable phase of matter where Maxwell electrodynamics is replaced by axion electrodynamics: a 3D topological insulator. 
\end{abstract}
\maketitle

\begin{figure}
\centering
\includegraphics[width=8.cm ]{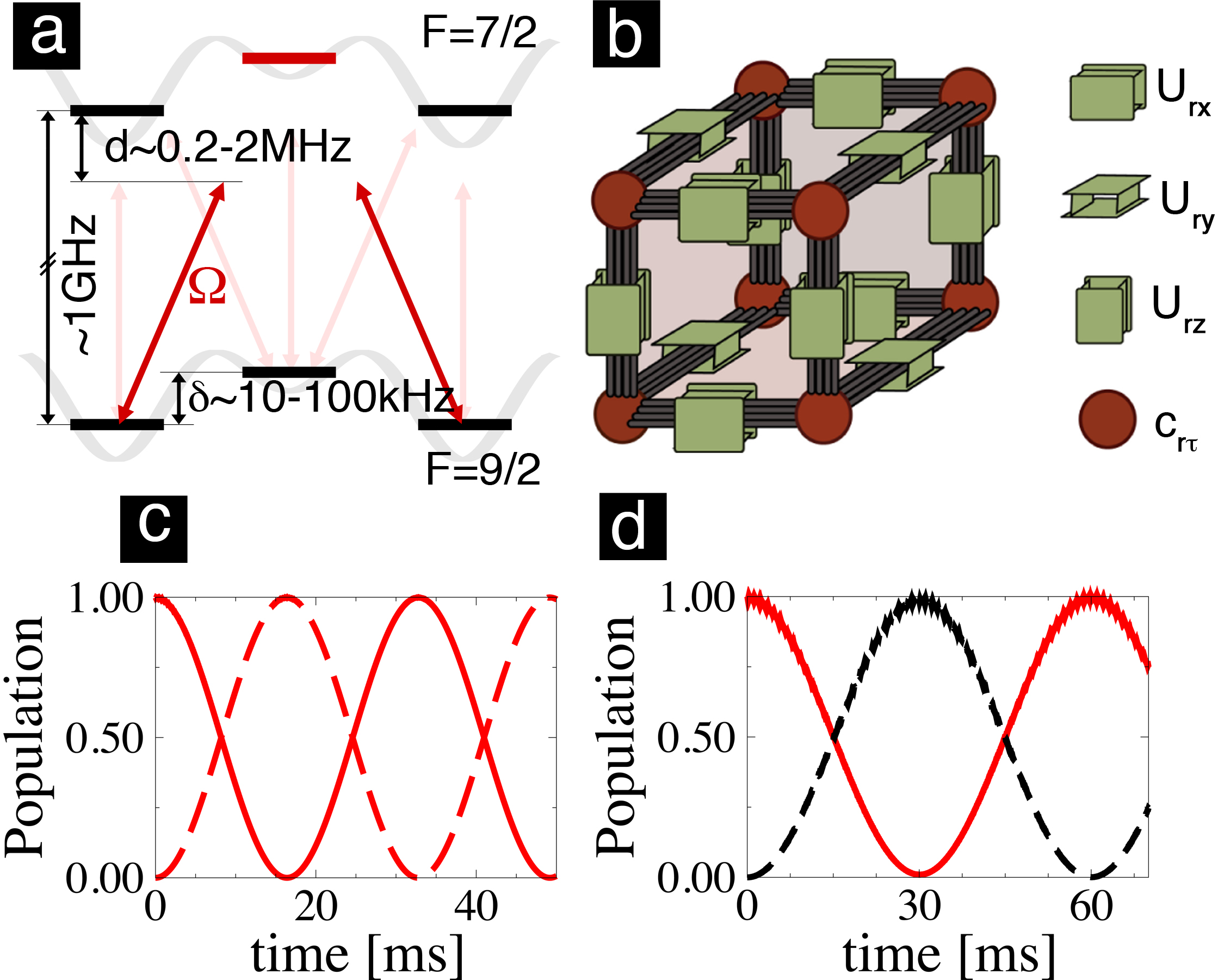}
\caption{ 
a) Superlattice potential (grey lines) trapping $^{40}$K atoms in the main and secondary  minima.  The hopping between  $F=9/2$ levels (in black) is laser-assisted via an intermediate $F=7/2$ state (in red). The  coupling, detuned by $d+\delta$, is induced by an off-resonant Raman transition with Rabi frequency $\Omega$. b) Scheme of the four states of the $F=9/2$ manifold (red vertices), connected by laser-induced hoppings (green boxes). c) Time-evolution of the pop-
ulations of the neighbouring hyperfine levels.
The solid (dashed) line is used for site $i$ $(i+1)$; the red (black) line
is used for $m_F=9/2$ $(m_F = 7/2)$. A clear spin-preserving
Rabi oscillation between neighboring sites is shown. The
numerical simulation is an exact Runge-Kutta time-evolution of the
complete model involving all the couplings and the levels in Fig. ~\ref{fig1}a.
d) The same as before for a spin-ßipping hopping. Notice the
need for a superlattice staggering (10-20 kHz) in order to avoid on-
site spin-ßipping. Exact time-evolution shows oscillations between
neighboring sites with a different spin.
}
\label{fig1}
\end{figure}

 The formulation of relativistic fermions in lattice gauge theories (LGTs)~\cite{kogut_rmp}  
is hampered by the fundamental problem of species doubling~\cite{karsten},  namely, the rise of
spurious fermions  that modify the physics at long wavelengths. To prevent the abundance of fermion doublers, a suitable tailoring of their  masses is required, leading to the so-called Wilson fermions~\cite{wilson}. 
A different, but also fundamental,  hindrance in high-energy physics is the strong CP problem, more precisely, the lack of  experiments confirming the charge-parity violation in strong interactions \cite{strong_CP1}.  To reconcile theory and experiment, the Peccei-Quinn mechanism postulates a new particle, the axion~\cite{axion1}, whose detection still remains elusive. Interestingly enough, these two seemingly unrelated problems turn out to be closely connected. Indeed, Wilson fermions with an inverted mass give rise  to a certain axion background~\cite{theta_zhang,axion_ti_zhang}. In this Letter, we suggest to exploit this connection to explore axion electrodynamics~\cite{axion_wilczek}
in a tabletop experiment of ultracold atoms. The exquisite and genuine 
control over quantum systems inherent to ultracold-atom technologies~\cite{ol_bloch} allows us  to propose the implementation Wilson fermions in optical superlattices (Fig.~\ref{fig1}a). This experiment
would trace a promising route to design, control, and probe the rich physics of axions.   As a first step,  inverting the Wilson  mass via laser-assisted tunneling  enables us to realize a background axion field $\theta = \pi$, which corresponds to a new class of unconventional states of matter: 3D topological insulators (TIs)~\cite{TI_review,3dti_fu_kane_mele,theta_zhang,axion_ti_zhang}. These gapped phases have conducting edges  protected by  topological order, but respect time-reversal symmetry. We show that our proposal constitutes the first fully-controllable quantum simulator (QS)~\cite{qs_feynman} of 3D TIs preserving a general anti-unitary symmetry. Besides, the space- and time-dependence of the axion field  can   be experimentally tailored   to test    the magnetoelectric~\cite{theta_zhang},  Witten~\cite{axion_ti_zhang}, or  Wormhole effects~\cite{wormhole_franz}. We focus instead on a fractional magnetic capacitor, whose implementation and  detection are better suited for optical-lattice techniques. Along this route, several intermediate, but  interesting, phenomena can be observed in our QS:
{\it i)} 3+1 massless Dirac fermions; {\it ii)} 3+1 massive Dirac fermions; and  {\it iii)} Wilson fermions.


We consider  a $^{40}$K Fermi  gas   in an optical superlattice~\cite{ol_bloch}, focusing in the Zeeman sublevels (i.e. spins) of the $F=\tiny{\frac{9}{2}}$  hyperfine manifold. Below,  we show that laser-assisted tunneling methods lead to the effective Hamiltonian ($\hbar=1$)
\begin{equation}
\label{initial_ham}
H_{\text{eff}}=\sum_{\boldsymbol{r}\boldsymbol{\nu}}\sum_{\tau\tau'}t_{\nu}c^{\dagger}_{\boldsymbol{r}+\boldsymbol{\nu}\tau'}[U_{\boldsymbol{r}\boldsymbol{\nu}}]_{\tau'\tau}c_{\boldsymbol{r}\tau}+\text{H.c.},
\end{equation}
where  $c^{\dagger}_{\boldsymbol{r}\tau}(c_{\boldsymbol{r}\tau})$ creates (annihilates) a fermion with spin $\tau$ at  site $\boldsymbol{r}=m\hat{\boldsymbol{x}}+n\hat{\boldsymbol{y}}+l\hat{\boldsymbol{z}}$ with $m,n,l\in\{1...L\}$, and
$t_{\nu}\sim 0.1$-$1$kHz is  the tunneling strength. Here, $U_{\boldsymbol{r} \boldsymbol{\nu}}$ are operators dressing the hopping from $\boldsymbol{r}\to\boldsymbol{r}+\boldsymbol{\nu}$, $\boldsymbol{\nu}\in\{\hat{\boldsymbol{x}},\hat{\boldsymbol{y}},\hat{\boldsymbol{z}}\}$, and we use gaussian units. These  operators   usually  rely on spin-dependent optical  lattices~\cite{gauge_jaksch}. We use instead  spin-independent bichromatic superlattices (Fig.~\ref{fig1}a), which trap  all  levels from  the $F=\{\tiny{\frac{7}{2}},\tiny{\frac{9}{2}}\}$ manifolds, and  allow for lifetimes $\tau_{\text{l}}\sim 1\text{s}$. The optical potential $V(\boldsymbol r) = V_0 \sum_{\nu} [ \cos^2 (\pi r_{\nu}) +  \cos^2 (2\pi r_{\nu})]$, where $V_0\sim50$-$150$kHz  and we have set the lattice spacing to 1. This  yields  a  cubic superlattice with atoms trapped in the  minima at zero energy (i.e. sites), and  secondary minima at  $\Delta E_{\text{sl}}\sim 50$-$100$ kHz (i.e. links). The hopping  $U_{\boldsymbol{r}\boldsymbol{\nu}}$ between  $F=\tiny{\frac{9}{2}}$  atoms  in neighboring sites is mediated by a Raman transition to a $F=\tiny{\frac{7}{2}}$ ``bus''  level  in the intermediate link (Fig.~\ref{fig1}a). 

Let us consider two states $\ket{\alpha, \zeta},\ket{\alpha', \zeta'}$, where $\alpha,\alpha '$ label the Zeeman sublevels of the $F=\{\frac{7}{2},\frac{9}{2}\}$ manifolds, whereas $\zeta,\zeta'$ label the band index and center-of-mass coordinates spanning the lattice sites and links. A two-photon process ($\omega_i$ and $\mathbf p_i$ are the frequency and momentum of the $i$-th photon), after eliminating an excited level, couples these  states 
\begin{equation}
H_{\text{L}}=\sum_{\alpha,\alpha'}\sum_{\zeta,\zeta'}\tilde{\Omega}^{\alpha' \zeta'}_{ \alpha \zeta} c^{\dagger}_{\zeta'\alpha'}c_{\zeta\alpha}+\text{H.c.},\hspace{1.5ex} \tilde{\Omega}^{\alpha' \zeta'}_{ \alpha \zeta} = S_{\zeta'\zeta}  \Omega_{\alpha' \alpha}  \ee^{-\ii \omega t}
\label{eq:shortRaman}
\end{equation}
where $c^{\dagger}_{\zeta\alpha}(c_{\zeta\alpha})$ creates (annihilates) a fermion in $\ket{\alpha, \zeta}$,  $S_{\zeta'\zeta} $ is the overlap between Wannier wavefunctions in presence of a momentum transfer $\mathbf p_1 - \mathbf p_2$: $S_{\zeta'\zeta} = \bra{\zeta'} e^{-i (\mathbf p_2 - \mathbf p_1) \cdot \mathbf x} \ket{\zeta}$, and $\omega = \omega_1 - \omega_2$. The remaining part of the coupling depends on the light polarization  and the atomic internal structure. Thus, the formula factors out the contribution of the center-of-mass wavefunction and of the internal degrees of freedom.
If the transferred momentum is large, we  increase the overlap factor $S_{\zeta'\zeta}$ between neighboring  sites and links, and thus the hopping. Besides,  if  lasers are far-detuned from this transition, we can adiabatically eliminate the ``bus'' level,  realizing a  four-photon coupling that leads to the single-band Hamiltonian in
 Eq.~\eqref{initial_ham}. Here, the hopping strengths scale as ${t}_{\nu}\sim|\Omega|^2/d$, where $d\sim0.2-2$MHz, and $\Omega\sim10$-$100$kHz is the Rabi frequency (Fig.~\ref{fig1}a). Due to the large Zeeman shift $\Delta E_z / B \sim 0.3$ MHz/G, we can independently implement each  matrix element of $U_{\boldsymbol{r} \boldsymbol{\nu}}$ eliminating a different  ``bus'' level. A careful analysis shows that the contributions of other bands and spurious on-site couplings can be neglected~\cite{Mazza}. In Figs.~\ref{fig1}c-d, we confirm this   for  two  schemes with the necessary ingredients:  spin-preserving and spin-flipping hoppings. We use 4 spin components, and design the hopping in terms of Pauli matrices (Fig.~\ref{fig1}b), $U_{\boldsymbol{r}\boldsymbol{\nu}}=\ee^{-\ii\phi_{\nu}\alpha_{\nu}}$, where $\alpha_{\nu}=\sigma_z\otimes\sigma_{\nu}$ and  $\phi_{\nu}\in\mathbb{R}$. Such block-structure   
allows the implementation of $U_{\boldsymbol{r}\boldsymbol{\nu}}$ in parallel for each spin pair, thus reducing the experimental intricacies.
 The diagonal tunneling  can be directly implemented (see Fig~\ref{fig1}c). The spin-flipping  hopping requires the even/odd  sites  to be staggered with $\Delta E_{\text{st}}\sim$10-20kHz, but  is also efficient  (Fig.~\ref{fig1}d). This scheme, originally developed for spin-1 bosons with  3-body interactions~\cite{Mazza}, leads to the interaction picture Hamiltonian in Eq.~\eqref{initial_ham}
when the  scattering is switched off by Feschbach resonances~\cite{feschbach}. 

Remarkably enough, starting from  this ultracold gas of non-relativistic  atoms, there are certain regimes where the emergent quasiparticles become ultra-relativistic fermions. We explore such avenues for a translationally-invariant system,  where the bulk energy bands come in degenerate pairs  $
E_{\boldsymbol{k}\pm}=\sum_{\nu}2t_{\nu}\cos k_{\nu}\cos\phi_{\nu}\pm2(\sum_{\nu}t_{\nu}^2\sin^2 k_{\nu}\sin^2\phi_{\nu})^{1/2}$, and $\boldsymbol{k}\in[-\pi,\pi]^3$ lies in the Brillouin zone. In the $\pi$-flux regime $  \phi_{\nu}=\pi/2$,  atoms wandering around plaquettes take on an overall minus sign, and the bands touch at  $N_{\text{D}}=8$ different points $
\boldsymbol{\Lambda}_{\boldsymbol{d}}\in\left\{\left(d_x{\pi},d_y{\pi},d_z{\pi}\right):\hspace{0.5ex} d_x,d_y,d_z=0,1\right\}$. Around them,  low-energy excitations display a relativistic dispersion
$E(\boldsymbol{p})\approx\pm (c_x^2p_x^2+c_y^2p_y^2+c_z^2p_z^2)^{1/2},$
where $c_{\nu}=2 t_{\nu}$ is the effective  speed of light, and $\textbf{p}=\textbf{k}-\boldsymbol{\Lambda}_{\boldsymbol{d}}$. Indeed,  imposing an ultraviolet  cutoff $|{p}_{\nu}|\leq 1/2\lambda_c$, the  effective field theory is
\begin{equation}
\label{dirac_ham}
H_{\text{eff}}^{\boldsymbol{d}}=\int_{\lambda_c} \text{d}^3r\Psi^{\dagger}(\textbf{r})H_{\text{D}}^{\boldsymbol{d}}\Psi(\textbf{r}), \hspace{1ex}H_{\text{D}}^{\boldsymbol{d}}=\sum_{\nu}c_{\nu}{\alpha}^{\boldsymbol{d}}_{\nu}p_{\nu},
\end{equation} 
where $\Psi(\textbf{r})=(c_1(\textbf{r}),c_2(\textbf{r}),c_3(\textbf{r}),c_4(\textbf{r}))^t$ is the  field operator, $\alpha^{\boldsymbol{d}}_{\nu}=(-1)^{d_{\nu}}\alpha_{\nu}$, and $ p_{\nu}=-\ii\partial/\partial r_{\nu}$ is the momentum.  The chosen hoppings induce a Clifford algebra $\{\alpha^{\boldsymbol{d}}_{\nu},{\alpha}^{\boldsymbol{d}}_{\mu}\}=2\delta_{\nu\mu}$, and  Eq.~\eqref{dirac_ham} yields a physical realization of the so-called {\it naive Dirac fermions} in LGT~\cite{karsten}. In our scheme,  fermion doubling leads to  an {\it even} number of species  which, in contrast to the  artificial doublers in LGT,   correspond to physical and observable flavors. We remark that each of them has a different chirality   $\gamma_5^{\boldsymbol{d}}=Q_5^{\boldsymbol{d}}\gamma_5$, where $\gamma_5=\sigma_z\otimes\mathbb{I}$, and $Q_5^{\boldsymbol{d}}=(-1)^{d_x+d_y+d_z}$ is the axial charge. Chiral symmetry, which plays a fundamental role in  the standard model  classifying right/left-handed particles $\gamma_5\Psi=\pm\Psi$, cannot be incorporated  to the lattice globally.

Let us stress that we are not limited to the massless limit, but can also explore a massive regime with  $H^{\boldsymbol{d}}_{\text{eff}}+H_{m}$, where  $H_{m}=\int \text{d}^3r\Psi({\boldsymbol{r}})^{\dagger}mc^2\beta\Psi({\boldsymbol{r})}$, and $\beta=\sigma_x\otimes\mathbb{I}$. Since no momentum transfer is required, this  term can be engineered via on-site microwave Raman transitions after adiabatic elimination of the   $F=\frac{7}{2}$ manifold, where $mc^2$ is the Raman strength.

In LGT,  Wilson envisaged a method to  decouple  the doublers from a single Dirac fermion~\cite{wilson}.  Our versatile  approach allows us to  realize his idea by combining  the previous ingredients with  additional tunnelings $\tilde{U}_{\boldsymbol{r}\boldsymbol{\nu}}=-\ii\ee^{-\ii{\varphi}_{\nu}\beta}$. For  $\varphi_{\nu}=\pi/2$, the  full Hamiltonian in Eq.~\eqref{dirac_ham} turns into
\begin{equation}
\label{wilson_mass}
H_{\text{eff}}^{\boldsymbol{d}}=\sum_{\nu}c_{\nu}{\alpha}^{\boldsymbol{d}}_{\nu}p_{\nu}+m_{\boldsymbol{d}}c^2\beta,\hspace{1ex} m_{\boldsymbol{d}}=m-\sum_{\nu}(-1)^{d_{\nu}}m_{\nu},
\end{equation} 
where   $m_{\nu}c^2=2\tilde{t}_{\nu}$  depends on the assisted-hopping strength, and thus on the laser power.  We highlight that the assisted hopping strengths scale as $\tilde{t}_{\nu}\sim|\Omega|^2/d$. Since $|\Omega|^2$ is proportional to the laser intensities, the tunneling strength, and thus the masses, are controlled by the beams power. We finally stress that since $t_{\nu},\tilde t_{\nu}\sim 0.1$-$1$ kHz, the temperature requirements of this proposal are of the same order as those faced in the context of quantum magnetism in optical lattices, currently at the forefront of experimental research. Setting $mc^2=2(\tilde{t}_x+\tilde{t}_y+\tilde{t}_z)$, doublers become very massive and decouple from the massless fermion at the center of the Brillouin zone $\boldsymbol{\Lambda}_{\boldsymbol{0}}=\boldsymbol{0}$. At the expense of breaking chiral symmetry $[H_{\text{D}}^{\boldsymbol{d}},\gamma_5]\neq0$, we have a QS of {\it Wilson fermions} invariant under the anti-unitary operator $\mathcal{U}_a=(\ii\mathbb{I}\otimes\sigma_y)\mathcal{K}$, where $\mathcal{K}$ is  complex conjugation. 

\begin{figure}
\includegraphics[width=8.cm]{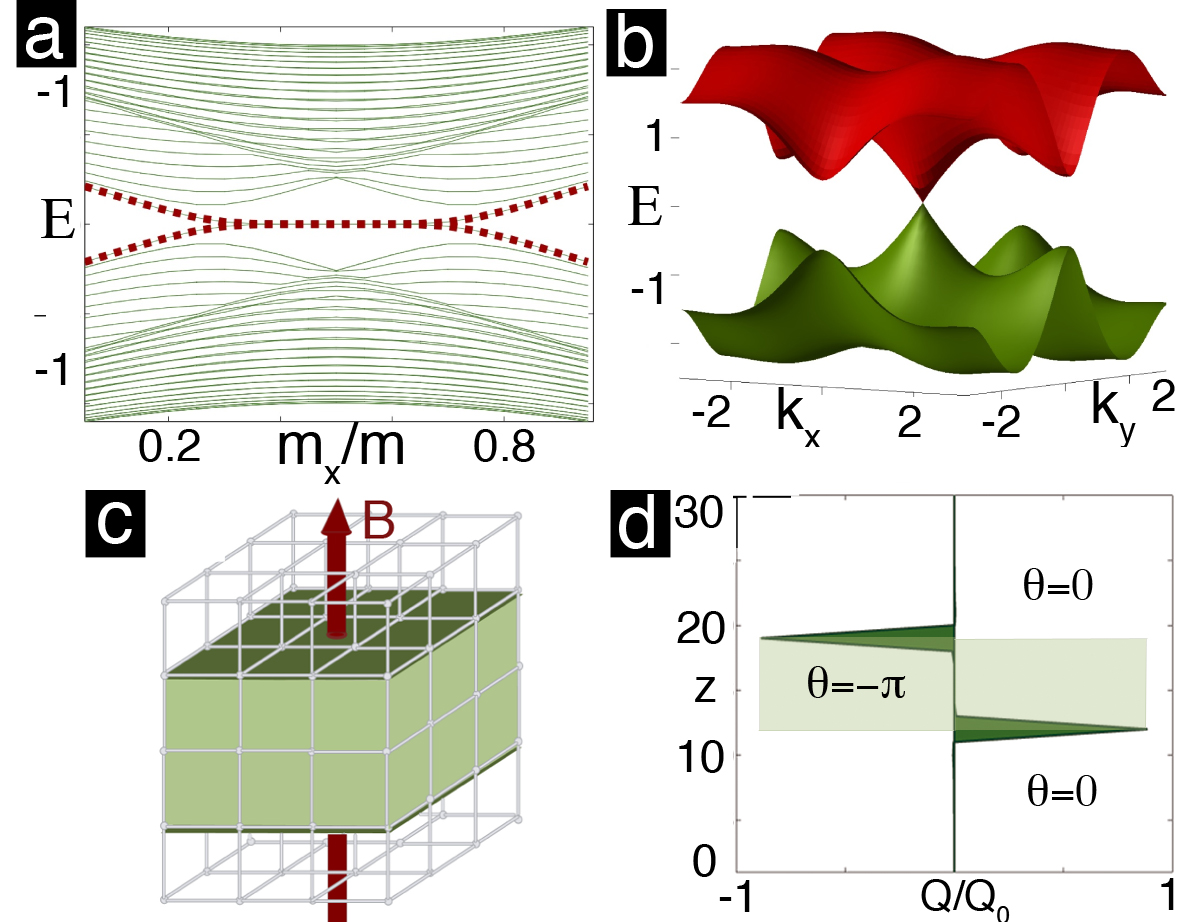}
\caption{
a) In-gap zero-energy modes (dashed red lines) for $\boldsymbol{q}=(k_x,k_y)=\boldsymbol{0}$,        $m/4\leq m_x\leq3m/4$, and $m_y=m/2,m_z=m/4$, for a lattice with $N=40^3$ sites and open boundaries at $z=0,L$.
b)  Boundary massless Dirac fermion at $z=0$,  $\boldsymbol{q}=\boldsymbol{0}$, and $m_x=m_y=m_z=m/2$.
c) Scheme for a fractional magnetic capacitor consisting of an axion well:  $\theta(\boldsymbol{r},t)=-{\pi}$ if $z\in[z_{\text{l}},z_{\text{r}}]$, and $\theta(\boldsymbol{r},t)=0$ elsewhere, pierced by a magnetic field $\boldsymbol{B}=B\hat{\boldsymbol{z}}$. This is  designed by tuning $m_{x}=m_y=m_z/2=m/4$ globally, whereas $\tilde{m}\gg m$ is only applied to $z_{\text{l}}<z<z_{\text{r}}$.
d) Accumulated charge on the ``plates'' of the capacitor, for a lattice of $N=30^3$ sites, $m_x=m_y=m_z/2=m/4$, $\tilde{m}=10m$ (leading to $\theta=-\pi$ for $12<z<18$), and flux $\phi/\phi_0=2\pi/15$.
}
\label{fig2}
\end{figure}

In an effort to preserve chiral symmetry, {\it  domain-wall fermions} are introduced in 4+1 dimensional LGTs~\cite{kaplan}, whose lower-dimensional descendants are the so-called topological insulators~\cite{TI_review,3dti_fu_kane_mele,theta_zhang,axion_ti_zhang}. These holographic phases have an insulating  bulk and metallic boundaries where topologically-protected midgap Dirac fermions reside. We can  realize these phases in experiments by inverting the sign of the Wilson mass through the laser intensity. We study the effect of  mass anisotropy on a lattice with open z-boundaries,  which leads to the energy spectrum in Fig.~\ref{fig2}a.  For a critical anisotropy $m_x^{c}=\frac{1}{4}m$, a   mass inversion occurs and some levels  leave the bulk bands to become zero-energy  states; whereas for $m_x^{c}=\frac{3}{4}m$ the midgap states fuse back into the continuum. In Fig.~\ref{fig2}b, the associated energy band displays a solitary 2+1 massless Dirac fermion exponentially localized around $z=0$, whereas its doubler appears at  $z=L$. In other words, there are two distant surfaces with an {\it odd} number of massless fermions, an unambiguous signal of  a strong TI that defies the doubling.

\begin{figure}[]
\centering
\includegraphics[ width=8.50 cm]{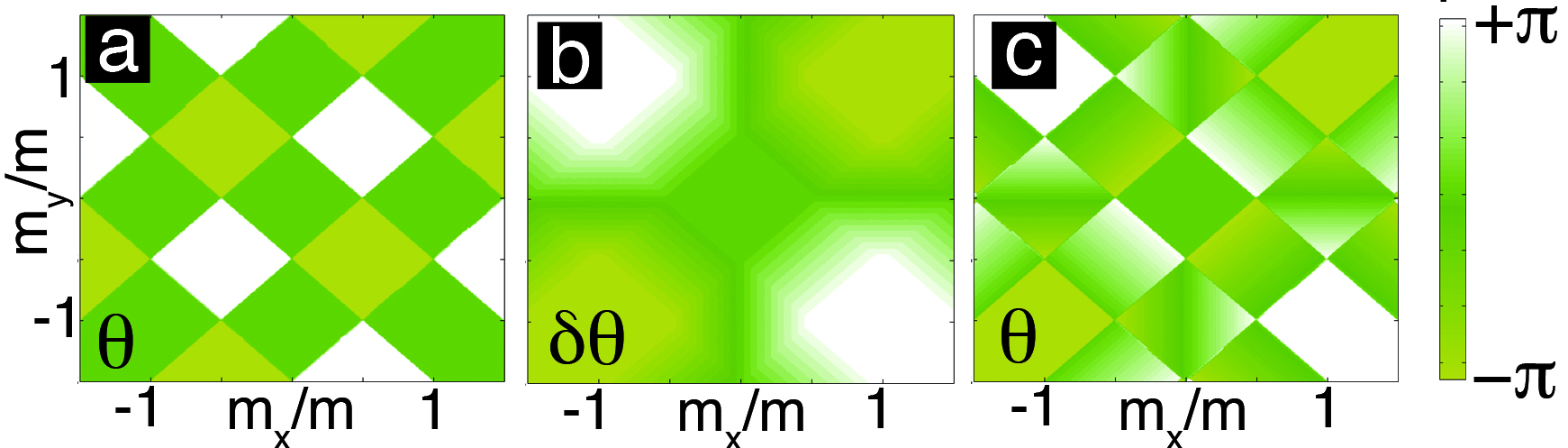} 
\caption{ a) Axion index as a function of the masses $m_y/m,m_x/m$, and setting $ m_z=m/2$. In the $\mathcal{U}_a$ invariant regime, only fixed values of the axion $\theta=\{0,\pi\}$ are allowed.  b) Perturbations to the axion term  $\delta\theta$ in the $\mathcal{U}_a$-breaking regime. d) Total axion term $\theta$ in the $\mathcal{U}_a$-breaking regime. }
\label{fig3}
\end{figure}

TIs encode a   topological order that  has dramatic consequences on their  response to electromagnetic fields~\cite{axion_wilczek, theta_zhang,axion_ti_zhang}. We explore these effects by synthesizing artificial $\boldsymbol{E},\boldsymbol{B}$ fields~\cite{gauge_jaksch}, subjected to a modified  Gauss law
$
\nabla\cdot \boldsymbol{E}=4\pi\rho-(e^2/\pi c)\nabla\theta\cdot\boldsymbol{B}
$, where $\rho$ is the atomic  density. For small  masses~\cite{theta_zhang,axion_ti_zhang}, and after a spin rotation,  the corresponding Bloch states lead to $\theta=-\frac{\pi}{2}\text{sgn}(c_xc_yc_z)\sum_{\boldsymbol{d}}Q_5^{\boldsymbol{d}}\text{sgn}(m_{\boldsymbol{d}})$, which only depends on the axial charge and the sign of the Wilson mass of  each bulk Dirac fermion.
 In Fig.~\ref{fig3}a, one  observes that the strong TIs correspond to a non-vanishing axion  field $\theta=\pi \text{ mod}(2\pi)$. Besides, this value is robust with respect to small errors in the implementation of the assisted hopping $c_{\nu}=c_{\nu}+\delta c_{\nu}$.
 
  It is important to notice that the anti-unitary symmetry  fixes $\theta=\{0,\pi\}$~\cite{theta_zhang}. However, it is possible to go beyond this scenario~\cite{dyn_axion}, and examine further possibilities lying in our proposal, by introducing a complex mass $m\beta+\ii \tilde{m}\beta\gamma_5$, leading to an arbitrary axion  $\theta\in[0,2\pi)$. Such a term, which  breaks  $\mathcal{U}_a$,  is obtained  via  Raman transitions,  and in  the isotropic limit leads to
\begin{equation}
\label{total_axion_term}
\theta=-\frac{\pi}{2}\sum_{\boldsymbol{d}}Q_5^{\boldsymbol{d}}\text{sgn}(m_{\boldsymbol{d}})-\sum_{\boldsymbol{d}}Q_5^{\boldsymbol{d}}m_{\boldsymbol{d}}\text{tan}^{-1}\left(\frac{\tilde{m}}{m_{\boldsymbol{d}}}\right),
\end{equation} 
which contains a perturbation $\delta\theta=-\sum_{\boldsymbol{d}}Q_5^{\boldsymbol{d}}m_{\boldsymbol{d}}\text{tan}^{-1}(\tilde{m}/m_{\boldsymbol{d}})$. Note that for $\tilde{m}\to0$, $\delta\theta\to 0$, we recover the previous result. Conversely, when  $\tilde{m}\gg|m_{\boldsymbol{d}}|$, the transparent expression  $\delta\theta=-\frac{\pi}{2}\sum_{\boldsymbol{d}}Q_5^{\boldsymbol{d}}|m_{\boldsymbol{d}}|$ is obtained (Fig.~\ref{fig3}b). This perturbation takes on any possible value $\delta\theta\in[-\pi,\pi)$ leading to the total axion  of Fig.~\ref{fig3}c. Therefore, our atomic gas constitutes a tunable axion medium where the time- and space-dependence of $\theta(\boldsymbol{r},t)$ can be externally adjusted by tailoring the focusing width and intensity of the Raman lasers $\tilde{m}(\boldsymbol{r},t)$. We can now explore an exotic consequence of axion electrodynamics: the fractional magnetic capacitor.
Engineering the axion medium in Fig.~\ref{fig2}c by focusing a strong Raman laser onto an inner region,  confines the axion  $\theta=-\pi$  to $z_{\text{l}}<z<z_{\text{r}}$, and minimizes the effects of the external trapping potential present in experiments~\cite{stanescu}.  Besides, we can synthesize a magnetic field generalizing the techniques in~\cite{gauge_vortices_spielman} to a lattice~\cite{gauge_jaksch}. The field  is introduced via Peierls substitution $t_{\nu}\to t_{\nu}\ee^{-\ii\int_{\nu}\text{d}\boldsymbol{r}\boldsymbol{A}}$ (likewise for $\tilde{t}_{\nu}$)  in the Landau gauge $\boldsymbol{A}=-\phi y\hat{\boldsymbol{x}}$, where $\phi$ is the magnetic flux in units of the flux quantum. Setting $\phi=2\pi p/q$, with $p,q\in\mathbb{Z}$, allows us to diagonalize the  open-boundary Hamiltonian in terms of $4qL$-bands per momentum inside a reduced magnetic Brillouin zone $\boldsymbol{q}\in[-\pi,\pi)\times[-{\pi/q},{\pi/q})$, and then obtain the density $\rho(z)$ from the occupied eigenstates. Then, the charge per quantum flux is  $Q=2\pi l_{B}^2\rho(z)$, where $l_{B}=(c/eB)^{1/2}$ is the magnetic length.  Subtracting the value in the absence of a magnetic flux  leads  to Fig.~\ref{fig2}d, which  confirms the continuum prediction $Q=Q_0\delta(z-z_{\text{l}})-Q_0\delta(z-z_{\text{r}})$, where $Q_0=\frac{e}{2}$,  for an optical lattice  with  $N=30^3$ sites, and magnetic flux $\phi=2\pi/15$. In such  background, the modified Gauss law predicts an accumulation of fractional charge per flux quantum  at the boundaries. Therefore, our axion medium plays the role of an exotic fractional capacitor  whose boundaries act as conducting plates. In strong contrast to usual capacitors,  the charge stored  is fractional, and rather than the electric field, it is the magnetic field which triggers the effect. Let us recall that the $^{40}$K atoms are neutral, and thus this effect corresponds to an accumulation of atomic density rather than charge.

We have described a versatile QS capable of realizing 3+1 massless/massive Dirac fermions, Wilson fermions, and 3D topological insulators as an axion medium. However, we still need to address the detection of such effects.  
To distinguish between the different types of relativistic  bulk fermions, it suffices to measure the number of Dirac points at zero energy~\cite{comment}. This can be accomplished by measuring the atomic density as a function of the chemical potential close to zero energy~\cite{detection_df}; a standard technique that relies on absorption images of the Fermi gas  (i.e. the shadow projected by the atoms on a CCD camera) after the atoms have been released from the trap. Considerably more challenging is  the detection of the edge states characteristic of TIs. The ratio between the number of edge and bulk modes makes the direct detection by  density measurements inefficient. Therefore,  new but also more demanding methods have been proposed, such as the population of  edges modes with bosons~\cite{stanescu}, or the use of Raman spectroscopy and Bragg scattering~\cite{detection_edges} . In this Letter, we exploit the consequences of the axion medium to propose a density-based measurement. Let us remark that the accumulation of density in the fractional magnetic capacitor does not suffer from an unbalanced edge-bulk density ratio. In fact, an extensive number of atoms will accumulate on the capacitor plates, which can be detected by  phase-contrast imaging methods. These methods do not require the trap release, such as absorption imaging, but rather recover the atomic density {\it in situ} by measuring the phase shift of the off-resonant light diffracted by the Fermi gas. Accordingly, phase-contrast imaging is a non-destructive method that has already been implemented for bosons and fermions~\cite{dispersive_imaging}.

In this Letter, we have presented a  feasible scheme of  laser-assisted tunneling  in 3D optical lattices, which allows us to design, control, and 
probe Wilson fermions. This approach appears as a promising route towards the first fully-tunable realization of 3D TIs. We have shown that fractional magnetic capacitors, predicted by axion electrodynamics, can be produced and detected using techniques from optical lattices. Besides, since the axion dynamics  is tunable and each boundary  can be singled out, phenomena such as the magnetoelectric effect or the boundary fractional quantum Hall effect 
can also be pursued. Additionally, the interplay between topological order, disorder, and  strong  interactions can be investigated.



 A.B. and M.A.M.D. thank  MICINN  FIS2009-10061, 
CAM QUITEMAD, European FET-7 PICC, UCM-BS GICC-910758 and FPU. M.R. acknowledges  EC FP7/2007-2013 (247687, IP AQUTE).
N.G. thanks the F.R.S-F.N.R.S. M.L. acknowledges  MICINN 
(FIS2008-00784 and QOIT), EU (NAMEQUAM), ERC (QUAGATUA) and Humboldt Foundation. L.M. and M.R. thank J.I. Cirac and U. Schneider for fruitful discussions in conceiving the superlattice setup.

\end{document}